\documentclass[aps,prl,twocolumn,showpacs,superscriptaddress]{revtex4-1}
\usepackage{graphics}
\usepackage{amsmath}
\usepackage{amsfonts}
\usepackage{bm}
\usepackage{color}
\usepackage{bbm}

\newcommand{\nnabla}{\mathbf \nabla}

\newcommand{\req}[1]{Eq.~(\ref{#1})}
\newcommand{\reqs}[1]{Eqs.~(\ref{#1})}
\newcommand{\rref}[1]{(\ref{#1})}

\newcommand{\A}{\mathbf{A}}
\newcommand{\p}{\mathbf{p}}

\renewcommand{\r}{\mathbf{r}}

\newcommand{\beq}{\begin{equation}}
\newcommand{\eeq}{\end{equation}}
\newcommand{\be}{\begin{equation}}
\newcommand{\ee}{\end{equation}}
\newcommand{\beqa}{\begin{eqnarray}}
\newcommand{\eeqa}{\end{eqnarray}}
\newcommand{\bea}{\begin{eqnarray}}
\newcommand{\eea}{\end{eqnarray}}
\usepackage{amssymb}
\usepackage{array}
\usepackage{amsmath}
\usepackage{graphicx}\usepackage{graphics}
\usepackage{dcolumn}
\usepackage{bm}\usepackage{varioref}

\newcommand{\B}{\mathbf B}

\begin{document}

\title{Aharonov-Bohm Oscillations in Singly-Connected Disordered Conductors}

\author{I.~L.~Aleiner}
\affiliation{Department of Physics, Columbia University,
New York, NY 10027, USA}

\author{A.~V.~Andreev}
\affiliation{Department of Physics, University of Washington,
Seattle, Washington 98195, USA}

\author{V.~Vinokur}

\affiliation{Materials Science Division, Argonne National Laboratory,
  Argonne, Illinois 60439, USA}

\date{\today}

\begin{abstract}

We show that transport and thermodynamic properties of \emph{singly-connected} disordered conductors
exhibit quantum Aharonov - Bohm oscillations
with the total magnetic flux through the system.
The oscillations are associated with the interference contribution from a special class of electron trajectories  confined to
the surface of the sample.

\end{abstract}


\maketitle

{\em Introduction}--Quantum coherence of electron motion dramatically affects
low temperature physics in disordered
conductors. Anderson localization~\cite{Anderson58} is
the most profound phenomenon. Even in the metallic regime, where
quantum corrections are relatively small, they
give rise a number of dramatic
effects due to their extreme sensitivity to magnetic field and
inelastic processes \cite{AltshulerReview,LeeRamakrishnan85}.
Celebrated examples are universal conductance
fluctuations (UCF)~\cite{AltshulerUCF,LeeStoneUCF},
magnetoresitance in weak magnetic fields~\cite{LarkinNMR}, and
Aharonov-Bohm (AB) oscillations~\cite{SpivakAB} in
thin mesoscopic cylinders and nanorings~\cite{Sharvin,Webb}.
We show that quantum interference corrections give rise to a novel type of AB
oscillations that exist in finite \emph{singly-connected} conductors.
They originate from the boundary of the sample and are associated with a special type of diffusive
trajectories that graze the sample boundary.

\begin{figure}[h]
\includegraphics[width=0.9\columnwidth]{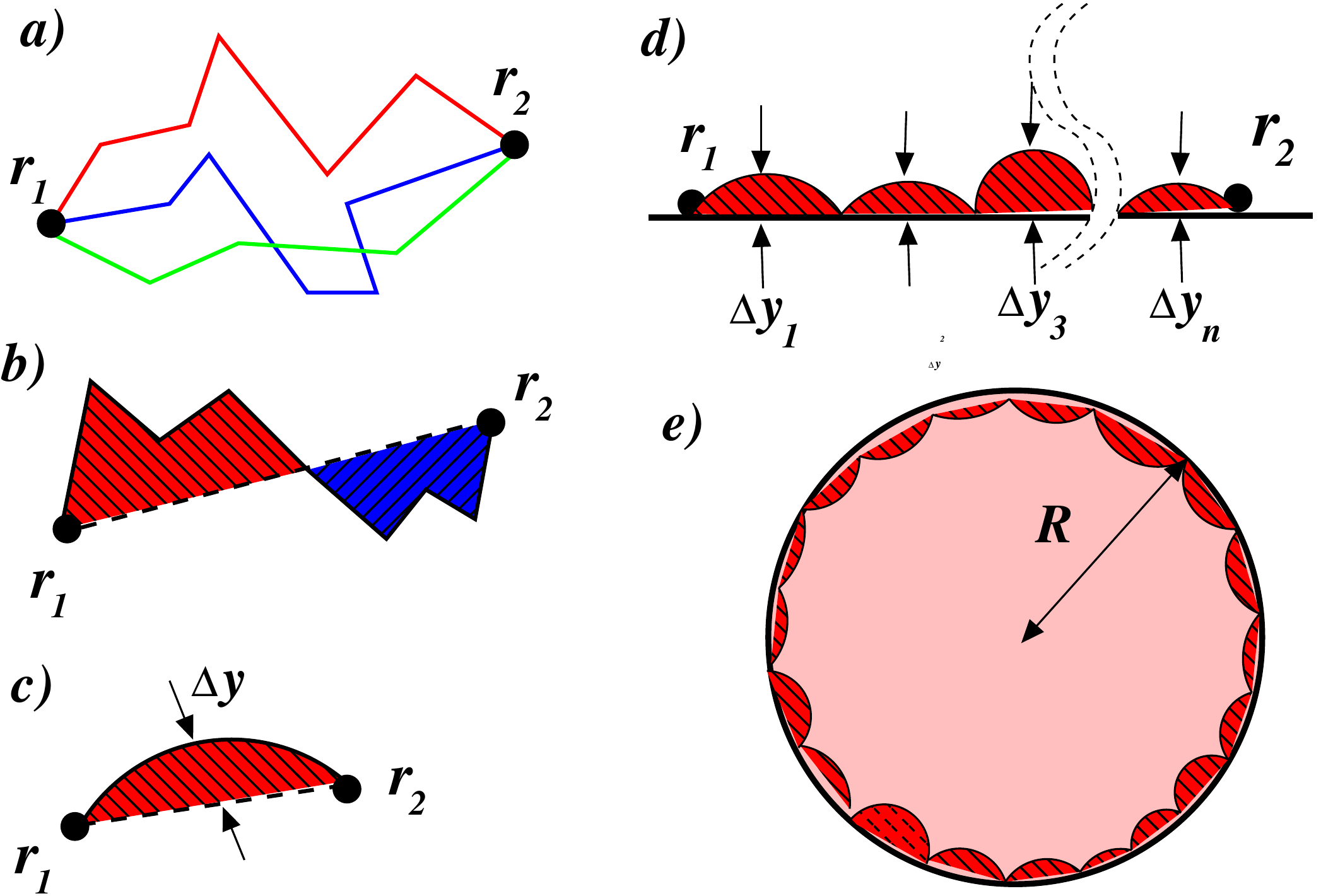}
\caption{a) Diffusive paths contributing to the interference corrections; b) Sketch of a directed area swept by path.
The blue and red regions give opposite in sign contributions; c)  Typical path leading to the gaussian decay
of the cooperon in the bulk; d) Typical path near the surface;  (e) Surface paths leading to AB oscillations.
Notice that the straight segment structure of diffusive paths a),b) is not shown in panels c)-e).}
\label{fig1}
\end{figure}

In the metallic regime quantum corrections may be understood semiclassically.
One can start with the classical motion of electrons
with  momentum $\p,\, |\p|\simeq p_F$ ($p_F$ is the Fermi momentum)
along diffusive trajectories (``paths'') consisting of segments of straight lines broken by impurities,
see Fig.~\ref{fig1} a). The  % \emph{quantum?}
 phase $\theta_l$ accumulated
during the motion along the $l$-th path is
\be
\theta_l(\B)=\frac{p_FL_l}{\hbar}+\frac{e}{c\hbar}\int_{l} d\r \cdot \A,
\label{phase}
\ee
where $L_l$ is the length of the path and the second term is the Aharonov-Bohm
phase due to the magnetic field, $\B=\nnabla \times \A$, and the integral is
evaluated along the classical path $l$.
Observables may be expressed in terms of the sum of quantum amplitudes taken
over all classical paths
connecting two points ($\r_1,\r_2$ in Fig.~\ref{fig1})
\be
\left|\sum_{l}\sqrt{W_l}e^{i\theta_l}\right|^2\!=\!
\sum_l W_l +2{\mathrm Re}\sum_{l\neq l'} \sqrt{W_lW_{l'}}
e^{i(\theta_l-\theta_{l'})},
\label{sum}
\ee where $W_l$ is the classical probability of path $l$.  The
first sum in the right hand side (rhs) of \req{sum} corresponds to the
classical probability of propagation from $\r_1$ to $\r_2$. The second
sum describes the quantum correction that arises from interference of
quantum amplitudes of different trajectories,
and is in general random due to strongly oscillating phase factors,
see the first term in the rhs of \req{phase}.
Thus, consideration of leading quantum corrections  reduces
to statistical analysis of these random terms, {\em e.g.} weak
localization correction is determined by its average whereas UCF are determined by its second cumulant, {\em etc.}.

One of the main quantities governing statistics of quantum corrections \cite{AltshulerReview} is the {\em cooperon}. It describes
interference of pairs of geometrically identical paths
traversed in opposite directions,
\be
C(\r_1,\r_2;\B)=\sum_l W_l e^{i[\theta_l(\B)-\theta_l(-\B)] }.
\label{Cooperon}
\ee
In the absence of magnetic field the phase factors above are equal to unity,
and the cooperon is given by the probability of classical diffusion between points $\r, \r^\prime$.
In a finite magnetic field the phase factors become random due to accumulation of the Aharonov-Bohm flux through oriented areas
swept by diffusive paths, see Fig.~\ref{fig1} b). It
is well known \cite{AltshulerReview} (more explanation will be given shortly) that
this randomness suppresses the cooperon at
distances larger than the magnetic length,  $l_B=\sqrt{\hbar c/e B}$,
so that for an infinite (translationally invariant)
system $C(\r_1,\r_2;\B) \simeq \exp\left[ 2ie/(\hbar c)\int \A d \r - \alpha_b (\r_1-\r_2)^2/l_B^2\right]$, where $\alpha_b$ a
is constant of order unity and the integral is taken along the chord connecting
$\r_1$ and $\r_2$.

As will be shown below, directed areas swept by electron trajectories near the boundary have smaller
randomness than those in the bulk. As a result, near the  boundary
the cooperon has a very different coordinate dependence dependence from that in the bulk;
 $C(\r_1,\r_2;\B) \simeq  \exp\left[\int_s \left(2ie/(\hbar c)\A d \r-\alpha_s | d \r| /l_B\right)\right]$.
Here $\alpha_s$ is a number of order unity, and the integration is taken along the  boundary. In contrast to the gaussian fall-off of the bulk contribution with the distance $|\r_1-r_2|$ the surface contribution falls off exponentially and thus becomes dominant at large distances.
Another important distinction is that the Aharonov-Bohm phase of the surface contribution is determined by geometry of the \emph{the sample boundary} rather than the shortest line connecting the end points.
Furthermore, the requirement that upon traversal of the perimeter of the sample, see Fig.~\ref{fig1} c),  the cooperon must remain unchanged leads to the conclusion
that the surface contribution $C(\r,\r;\B)$ is, in fact, an oscillating function of the magnetic flux through the sample, $\Phi$,
$C(\r,\r;\B)\simeq ...+ \exp\left(-\alpha_sp/l_B\right)\cos\left( 2\pi\Phi/\Phi_0\right)$, where $\Phi_0= 2\pi \hbar  c/2e$ is the superconducting flux quantum
and $p$ is the perimeter of the sample
(for 3-dimensional system $\Phi$ is calculated through the extremal crosssection
perpendicular to the magnetic field, and $p$ is the perimeter of such crosssection).
This means that the quantum interference corrections are predicted to \emph{oscillate} (rather than simply decay)
with the magnetic field even in singly-connected
geometries \cite{Bogacheck}, somewhat similar to the effects
in multiple connected geometries \cite{Sharvin,SpivakAB,LittleParks}.
We will discuss
possible experimental setups where these oscillations can be most readily observed.

{\em Qualitative discussion} -- To elucidate the difference between the surface and the bulk contribution to the interference corrections we now
discuss statistics  diffusive trajectories in more detail. Let us label each diffusive trajectory by $\r(\tau)$, $0\leq \tau \leq t$, where $t$ is the duration of the diffusive motion.  Then
summation over the paths $l$  can be re-written as
\[
\sum_lW_l\to \int d t P(t);\ P(t)\simeq\sum_{\r(\tau)}
\exp\left(-\frac{\int_0^t d\tau\dot{\r}^2}{4D}\right),
\]
where $D$ is the diffusion constant, and the summation (path integration) is performed over all trajectories $\r(\tau)$ in which the particle travels from $\r$ to $\r'$ in time $t$.  In this notation the bulk contribution to the Cooperon \rref{Cooperon} acquires the form
\be
C(\dots)=
e^{i\theta^{AB}_{\r_1\r_2}}\!
\int\! d t P(t) \left\langle\!
\exp\left(\!\frac{2iS\left\{\r(t)\right\}}{l^2_B}\right)\!\right\rangle,
\label{Cooperon1}
\ee
where the Aharonov-Bohm phase
$\theta^{AB}_{\r_1\r_2}=\frac{i2e}{\hbar c}
\int_{\r\r_2}d\r\cdot \A$ is calculated along the straight line connecting points $\r_1$  and $\r_2$,
$S\left\{\r(t)\right\}$ is the directed area of the surface confined by the path $\r(t)$ and the straight line connecting points $\r_1,\r_2$, see Fig.~\ref{fig1}, and
the averaging means
\be
\langle\dots \rangle=\frac{1}{P(t)}\sum_{\r(\tau)} \dots
\exp\left(-\frac{\int_0^t d\tau\dot{\r}^2}{4D}\right).
\ee

Some conclusions about  the statistical properties of $S$ can be easily drawn on
symmetry and dimensionality grounds. Consider for example a typical diffusive \emph{}path shown on
Fig.~\ref{fig1} c). Its probability
can be estimated as $\exp\left[-(\r_{12}^2+(\Delta y)^2)/(4Dt)\right]$ while
the directed area swept by it is $S\simeq 1/2 |r_{12}|\Delta y$.
Then, the averaged phase factor is given by
\be
\begin{split}
 \left\langle\!
\exp\left(\!\frac{2iS}{l^2_B}\right)
\!\right\rangle&=
\int_{-\infty}^\infty \frac{d\Delta y}{\dots} \exp\left(\!\frac{ir_{12}\Delta y}{l^2_B}\right)\exp\left(-\frac{(\Delta y)^2}{4Dt}\right)
\\
&=
\exp\left(-Dt\r_{12}^2/l_B^4\right)
\end{split}
\raisetag{9pt}
\label{Cooperon2}
\ee
(normalization factor is easily restored from $l_B\to \infty$ value).
Substituting this estimate into \req{Cooperon1}, we obtain
\[
C(\dots)\simeq \int dt \exp\left[-\r_{12}^2/(4Dt)\right]\exp\left(-Dt\r_{12}^2/l_B^4\right).
\]
The exponent in the integrand has a minimum at  $Dt \simeq l_B^2$ and at $|r_{12}|\gg l_B$
can be evaluated in the saddle point approximation with the result that in the \emph{interior of the system} the cooperon decays  rapidly at large distances,
\be
C(\r_{12};B)\simeq \exp\left(-\r_{12}^2/l_B^2\right).
\label{Cooperon3}
\ee
This rapid spatial decay of the bulk contribution arises from an \emph{unconstrained} summation
of a large number of contributions in \rref{Cooperon2}, which have random signs and nearly cancel each other.
The presence of a nearby boundary imposes a sharp geometrical constraint on the allowed paths,  which enhances the sum of rapidly oscillating contributions of different paths.
For instance, consider the same trajectory as in Fig.~\ref{fig1} c), but for points $\r_1,\r_2$ near the boundary. For such trajectories the geometric constraint imposed by the boundary amounts to confining the integration variable $y$ to the interval $y\in [0,\infty)$ resulting in
\[
\begin{split}
\dots
\left|\int_0^\infty dy \exp\left(\!\frac{ir_{12}\Delta y}{l^2_B}\right)\exp\left(-\frac{(\Delta y)^2}{4Dt}\right)\right|
 \simeq \frac{l^2_b}{|r_{12}|\sqrt{Dt} }
%\gg \exp\left(-Dt\r_{12}^2/l_B^4\right).
\end{split}
\]
Although this particular contribution decays merely as a power law at $\r_{12} \gg l_B$,
there are other oscillating contributions, which together  lead to an exponential decay. Indeed, consider a trajectory of the form of the ``skipping orbit''  shown on Fig.~\ref{fig1} d), which includes the $n$ reflections from the boundary.
Its probability is
$\exp\left(-\r_{12}^2/(4Dt)-n\sum_{j=1}^n(\Delta y_j)^2/(4Dt)\right)$, while the directed area swept it is $S\simeq 1/(2n) |r_{12}|\sum_{j=1}^n\Delta y_j$.
The estimate \rref{Cooperon2}, thus, immediately changes to
\[
  \left\langle\!
 \exp\left(\!\frac{2iS}{l^2_B}\right)
 \!\right\rangle =
  \prod_{j=1}^n\int\limits_{0}^\infty\frac{ d\Delta y_j}{\dots}
 \exp\left(\!\frac{ir_{12}\Delta y_j}{n l^2_B}-n \frac{(\Delta y_j)^2}{4Dt}\right).
\]
Estimating $n$ from the requirement that the contribution from both terms in the exponent be of the same order for the relevant value of
$\Delta y_j$, we find the optimal value of $n$
to be $n^3_* \simeq (Dtr_{12}^2)/l_B^4$.
The final value of the averaged phase factor becomes
\[
 \left\langle\!
 \exp\left(\!\frac{2iS}{l^2_B}\right)
 \!\right\rangle = \left(e^{-\alpha_s}\right)^{n_*}= \exp\left[-\alpha_s\left(\frac{Dtr_{12}^2}{l_B^4}\right)^{1/3}\right],
\]
where $\alpha_s$ is a coefficient of order unity (which will be found later) and $\mathrm{Re}\, \alpha_s>0$.
Substituting this estimate into \req{Cooperon1}, we find
\[
C(\dots)\simeq \int dt \exp\left[-\r_{12}^2/(4Dt)\right]\exp\left[-\alpha_s\left(\frac{Dtr_{12}^2}{l_B^4}\right)^{1/3}\right].
\]
The exponent in the integrand has a minimum at  $Dt \simeq l_Br_{12}$ and at $|r_{12}|\gg l_B$ we obtain the main qualitative result
\be
C_s(\r_{12};B)\simeq \exp\left(-\alpha_s|\r_{12}|/l_B\right),
\label{Cooperon4}
\ee
which significantly exceeds the bulk value of the cooperon \rref{Cooperon3}.
A similar enhancement of electron tunneling due to the surface effects
in the context of  an electron tunneling in magnetic field was
pointed out in Ref.~\cite{Shklovskii}. The similarity to our problem is
superficial as we consider the case of classically weak magnetic
field where diffusive trajectories of electrons are not bent
by the magnetic field. The latter enters only through the accumulation of the Aharonov-Bohm phase.

{\em Quantitative analysis of the cooperon} --
The qualitative picture above is borne out
by quantitative analysis. The cooperon represents the resolvent of
the modified diffusion equation \cite{AltshulerReview}
\begin{subequations}
\label{Cooperoneqn}
\be
\hbar D\left[
-i \nnabla - ({2 e}/{\hbar c}) \A(\r)
\right]^2
\chi_\beta(\r)=\epsilon_{\beta}\chi_\beta(\r),
\ee
supplemented with the boundary condition
\be
\left.\mathbf{n}\cdot\left[-i \nnabla - ({2 e}/{\hbar c}) \mathbf{A}(\mathbf{r})\right] \chi_\beta(\r)\right|_{\mathbf{r}\in \mathcal{B}}=0.
\ee
\end{subequations}
Here $\mathbf{n}$ is a vector normal to the boundary $\mathcal{B}$.
The vector potential $\A$ describes the effect of the Aharonov-Bohm phase accumulation and the boundary condition corresponds to the absence of current through the boundary. The eigenvalues $\epsilon_{\beta}$ are gauge invariant
and in many cases (see below) the physical effects are determined only by them.
The cooperon \rref{Cooperon} can be easily expressed as $C(\r_1,\r_2)=\sum_{\beta}
\chi_\beta(\r_1)\chi^*_\beta(\r_2)/\epsilon_\beta$.

\begin{figure}[h]
\includegraphics[width=0.8\columnwidth]{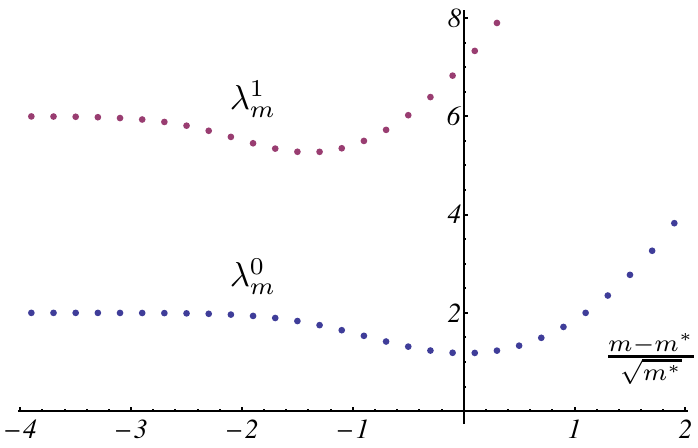}
\caption{Cooperon eigenvalues calculated for a disk geometry.
The drop near $m=m^*$ corresponds to the interference of
diffusion trajectories near the boundary shown in Fig.~\ref{fig1}.}
\label{fig2}
\end{figure}

To investigate the surface contribution to Aharonov-Bohm oscillations in a singly-connected
sample it is sufficient to use the simplest disk geometry  shown on Fig.~\ref{fig1} e).
Equations \rref{Cooperoneqn} are easily solved in the polar coordinates $r,\varphi$ in the
symmetric gauge $A_r=0,\ A_\varphi=-\frac{r B}{2}$.
The eigenstates in this case a labeled by two integers $\beta \to (n,m)$,  where
$m$ is the angular momentum, and the radial number  $n\geq 0$. The wave functions are
of the form $\chi=e^{i m\varphi}f_{m}^n(r/l_B)$.
The  eigenvalue $\epsilon_{m}^n=\lambda_{m}^n\hbar D/l_B^2$, and
the radial wave function $f_{m}^n(\rho)$ ($\rho=r/l_B$) obey
the dimensionless differential equation
\be
\left[-\frac{d^2}{d\rho^2}- \frac{1}{\rho}\frac{d}{d\rho}-2 m + \frac{m^2}{\rho^2} + \rho^2\right]f_{m}^n(\rho)=\lambda_{m}^nf_{m}^n(\rho).
\label{Landau-polar}
\ee
The Neumann  boundary condition $\left.\frac{df_{m}^n}{d\rho}\right|_{\rho=R/l_B}=0$ makes this problem different
from that for an electron in a magnetic field. The solution of \req{Landau-polar} that is regular at $\rho=0$ may be expressed in terms of the confluent hypergeometric function $\Phi(\alpha, \beta, z)$,
\[
f_{m}^n(\rho)=e^{-\rho^2/2}\rho^{|m|}
\Phi\left(\frac{|m|+1- m }{2}-\frac{\lambda_{m}^n}{4},|m|+1, \rho^2 \right).
\]
The eigenvalues $\lambda_{m}^n$ are found from the
boundary condition at $\rho=R/l_B$, for small angular momenta, $m\ll R/l_B$,
$\lambda_m^n=4n+2$ correspond to degenerate Landau levels, but
show significant deviations near the boundary.
For the two lowest Landau levels $\lambda_{m}^n$
are plotted in Fig.~\ref{fig2} for $\Phi/\Phi_0=25$.

As shown below, the magnitude of the Aharonov-Bohm oscillations is governed by the spectrum in the vicinity of the lower boundary,
$\lambda^*={\rm min}_m\lambda_{m}^0\approx 1.18$,
which is achieved at $m=m^*\approx \Phi/\Phi_0$.
Near the minimum the spectrum may be approximated
as
\be
\lambda_{0,m}\approx
\lambda^*\left(1+\frac{(m-m^*)^2}{\gamma^2 m^*}\right); \quad m^* \approx \frac{\Phi}{\Phi_0} \gg 1,
\label{minimum}
\ee
where $\gamma\approx 1.4 $.

To draw a connection to the boundary contribution
to the cooperon, see \req{Cooperon4},
we analyze the asymptotic behavior of the exact expression
\[
C(\r,\r')=\sum_{mn}\frac{e^{im(\phi-\phi')}f_{m}^n(r/l_B).
f_{m}^n(r'/l_B)}{\epsilon_{m}^n}
\]
At large distances, $|\phi-\phi'|m^* \gg 1 $ and $r \simeq r' \simeq m^* l_B$,
the exponential decay is determined by the pole closest to the real axis,
and we obtain
\[
C \simeq \exp\left[im^*(\phi-\phi')- \gamma\sqrt{m^*}|\phi-\phi'| \right],
\]
which corresponds to the edge physics discussed above.

{\em Physical manifestations} -- The surface contribution to the cooperon
gives rise to anomalous magnetic field behavior of the well known quantum interference corrections to thermodynamic and transport characteristics of disordered conductors. To be concrete,
we consider the simplest fluctuation correction to the thermodynamics of superconductor a in the normal state. The surface contribution
becomes especially well pronounced in the vicinity of the critical field $H_{c3}$~\cite{Saint-James}. The fluctuation correction to the free energy is connected to the eigenvalues \rref{Cooperoneqn} by~\cite{AltshulerReview}
\be
\delta F=T \sum_{k,\beta}\ln\left[\ln\left(\frac{T}{T_{c0}}\right)
+ \Psi\left(\frac{|k|}{2} + \frac{\epsilon_\beta}{4\pi T}\right) \right],
\label{fluctuation1}
\ee
where $\Psi(x)\equiv \psi(x+1/2)-\psi(1/2)$ with $\psi(x)$
being the digamma function.
The zeroth Matsubara frequency $k=0$, corresponds to the classical fluctuations,
whereas the quantum fluctuations are encoded into summation over all $k$.
At zero magnetic field ${\rm min}\, \epsilon_\beta=0$ and the correction \rref{fluctuation1} is meaningful only for $T>T_{c0}$, where the normal phase is stable.

At a finite magnetic field $\epsilon_\beta\geq \lambda_* \hbar D/l_B^2=\lambda_*{eDB}/{c}$. The normal phase is stable even at $T<T_{c0}$ for  $B>H_{c3}(T)$,
where surface superconductivity emerges \cite{Saint-James},
\be
\ln\frac{T_{c0}}{T}=\Psi\left(\lambda^*\frac{eDH_{c3}(T)}{4\pi c T}\right).
\label{Hc3}
\ee

The part of the free energy oscillating with the total flux $\Phi$ can be easily found analytically using \reqs{Landau-polar} -- \rref{minimum}.
Summing over $m$ with the aid of the Poisson summation formula, we obtain
 \be
\begin{split}
F_{os}&=T \sum_{k,n\neq 0}\int_{-1/2}^\infty \! dm e^{i2\pi mn}
\\
\times
&
\ln\left[\ln\left(\frac{T}{T_{c0}}\right)
+ \Psi\left(\frac{|k|}{2} + \frac{\lambda_0(m) e D B}{4\pi c T}\right) \right].
\end{split}
\label{fluctuation1}
\ee
At $m^*\gg 1$ the integral is determined by the branch cut of the logarithm. The branching points $m_b$ are determined by
$\lambda_0(m_b)=\lambda^* -2\pi cT|k|/(eDB)$. In the vicinity of $H_{c3}$ this condition simplifies to
\[
m_b=m^* \pm i \gamma\sqrt{m^*} \sqrt{\frac{B-H_{c3}(T)}{B} +\frac{2\pi |k| T}{T_c^0}\frac{H_c^3(0)}{B}},
%\label{brachingpoint}
\]
where we used \reqs{minimum} and \rref{Hc3}. After the integration along the branch cut and summation over the Matsubara frequencies $k$, we obtain
\be
\begin{split}
&-F_{os}=\sum_{n=1}^\infty  \frac{2 w_n\! \cos  2\pi n m^* }{n} =\sum_{n=1}^\infty \frac{2 w_n}{n} \cos \left(\frac{2\pi n \Phi}{\Phi_0}\right);
\\
&w_n=\sqrt{\frac{m_3}{m^*}}
\left(\frac{2 T_c B}{\pi H_{c3}(0)}\right)
e^{-\sqrt{\frac{m^*}{m_3}}}\Upsilon\left(
\sqrt{\frac{m^*}{m_3}}\frac{\pi T H_{c3}(0)}{2T_{c}B }
\right).
\end{split}
\label{Fosc}
\ee
Here the ``correlation momentum'' $m_3$ is determined by the proximity of the magnetic field to the value of $H_{c3}(T)$
\be
m_3=\frac{1}{\gamma^2}\frac{B}{B-H_{c3}}; \ \sqrt{\frac{m^*}{m_3}}
= \frac{\gamma}{2\pi} \frac{p}{l_B}\sqrt{\frac{B}{B-H_{c3}}}.
\ee
where $p=2\pi R$ is the perimeter of the sample. The last form remains valid for two-dimensional samples of non-circular shape.
The function $\Upsilon(x)\equiv x \coth x$ describes the crossover from the quantum ($x\ll 1$) to classical ($x\gg 1$)
fluctuation regime. In contrast to a bulk system the characteristic crossover scale depends on the length of the boundary, which
emphasizes the surface origin of the effect.

Equation \rref{Fosc} is the main illustrative result for the surface interference contribution we discussed.
Similar oscillations should appear  not only in
thermodynamic but also in transport properties of mesoscopic singly
connected devices, e.g. Aslamazov-Larkin corrections
to the conductance \cite{Galitski_Larkin2001} of
normal systems or  superconductor/normal metal hybrid structures,
somewhat similar to results \cite{Glazman_Hekking_Zyuzin92}
for thin cylinders.  Quantitative investigation of transport effects requires analysis of current redistribution near the edges of the sample and
will be reported elsewhere.
Oscillatory flux dependence of conductance of singly
connected wires and SNS junctions was recently reported in
Refs.~\onlinecite{DynesOSCMR,Kwok,Shahar,Markovic}. The observations of Ref.~\onlinecite{Shahar}
were interpreted in \cite{Pekker} in terms of formation of superconducting vortices inside the sample. We show here that the existence of AB oscillations in diffusive singly connected conductors is a much more general phenomenon which occurs even in the absence of superconductivity.

{\em In conclusion,} we identified a novel contribution to the magnetic field dependence of quantum interaction corrections in finite conductors. It arises from diffusive trajectories confined to the surface of the sample and gives rise to Aharonov-Bohm oscillations even in singly connected samples. In non-singly-connected samples or samples with holes or cavities,  AB oscillations will have multiple periods determined by the areas of extremal sections for each bounding surface.

We thank B. Spivak for useful discussions. Work at Columbia University was
 supported by the Simons foundation. Work at the University of Washington and  was supported by the U. S. Department of Energy
Basic Energy Science Program by the award number DE-FG02-07ER46452. Work at
 Argonne National Laboratory was
supported by the US Department of Energy, Office of Science, Materials Sciences and
Engineering Division (AA was partly supported at Argonne through the Materials Theory Institute).

\end{document}